\documentstyle[epsfig,axodraw]{aipproc}

\newcommand{\lsim}
{\mathrel{\raisebox{-.3em}{$\stackrel{\displaystyle <}{\sim}$}}}
\newcommand{\gsim}
{\mathrel{\raisebox{-.3em}{$\stackrel{\displaystyle >}{\sim}$}}}
\def\asymp#1%
{\mathrel{\raisebox{-.4em}{$\widetilde{\scriptstyle #1}$}}}

\def\Nequal#1%
{\mathrel{\raisebox{-.5em}{$\stackrel{=}{\scriptstyle\rm#1}$}}}
\newcommand{\dsl}[1]{\not \hspace{-0.7mm}#1}
\def\dsl{\mathpalette\make@slash}
\def\make@slash#1#2{\setbox\z@\hbox{$#1#2$}%
  \hbox to 0pt{\hss$#1/$\hss\kern-\wd0}\box0}

\begin{document}

\renewcommand{\thefootnote}{\fnsymbol{footnote}}
\thispagestyle{empty}
\strut\hfill ER/40685/948 \\
\strut\hfill UR-1609 \\
\vfill
\vspace*{1.5cm}
\begin{center}
{\Large\bf Electroweak Precision Physics \\[.3em] 
at \boldmath{$\mathrm{e^+e^-}$} Colliders with {\sc RacoonWW}
\footnote[2]{To appear in the {\it Proceedings of the 22nd annual MRST 
Conference on Theoretical High Energy Physics} (MRST 2000), 
Rochester, New York, May 8-9, 2000.}}
\\[1.4cm]
{\large A. Denner$^1$, S. Dittmaier$^2$, M. Roth$^3$,
and D. Wackeroth$^4$} \\[1.5em]
\parbox{10cm}{{\small 
$^1$Paul-Scherrer-Institut, CH-5232 Villigen PSI, Switzerland} 
\\ [0.5em]
{\small $^2$Universit\"at Bielefeld, D-33615 Bielefeld, 
Germany} \\ [0.5em]
{\small $^3$Universit\"at Leipzig, 
D-04109 Leipzig, Germany} \\ [0.5em]
{\small $^4$University of Rochester, 
Rochester, NY 14627, USA}}
\end{center}
\vspace*{2.5cm}
{\large\bf Abstract}\\[.2cm]
We present precise predictions for the processes ${\mathrm{e^+ e^- \to
WW}} \to 4f(\gamma)$ at LEP2 and future Linear-Collider (LC) energies
obtained with the Monte Carlo generator {\sc RacoonWW}.  The program
{\sc RacoonWW} includes the complete ${\cal O}(\alpha)$ electroweak
radiative corrections to ${\mathrm{e^+ e^- \to WW}} \to 4f$ in the
double-pole approximation (DPA).  While the virtual corrections are
treated in DPA, the calculation of the bremsstrahlung corrections is
based on the full lowest-order matrix elements to the processes
${\mathrm{e^+ e^-}} \to 4f+\gamma$.  This asymmetric treatment of
virtual and real photons requires a careful matching of the arising
infrared and collinear singularities.  We also take into account
higher-order initial-state photon radiation via the structure-function
method.  Here, we briefly describe the {\sc RacoonWW} approach, give
numerical results for the total W-pair production cross sections,
confront them with LEP2 data, and study the impact of the radiative
corrections on angular and W-invariant-mass distributions at LEP2 and
LC energies.

\clearpage

\setcounter{page}{1}
\renewcommand{\thefootnote}{\arabic{footnote}}

\title{Electroweak Precision Physics \\[.3em] 
at \boldmath{$\mathrm{e^+e^-}$} Colliders with {\sc RacoonWW}}

\author{A. Denner$^*$, S. Dittmaier$^{\dagger}$, M. Roth$^{\ddagger}$,
and D. Wackeroth$^{\P}$}
\address{$^*$Paul-Scherrer-Institut, CH-5232 Villigen PSI, Switzerland \\
$^{\dagger}$Theoretische Physik, Universit\"at Bielefeld, D-33615 Bielefeld, 
Germany \\
$^{\ddagger}$Institut f\"ur Theoretische Physik, Universit\"at Leipzig, 
D-04109 Leipzig, Germany \\
$^{\P}$Department of Physics and Astronomy, University of Rochester, 
Rochester, NY 14627, USA}

\maketitle

\begin{abstract}
We present precise predictions for the processes ${\mathrm{e^+ e^- \to
WW}} \to 4f(\gamma)$ at LEP2 and future Linear-Collider (LC) energies
obtained with the Monte Carlo generator {\sc RacoonWW}.  The program
{\sc RacoonWW} includes the complete ${\cal O}(\alpha)$ electroweak
radiative corrections to ${\mathrm{e^+ e^- \to WW}} \to 4f$ in the
double-pole approximation (DPA).  While the virtual corrections are
treated in DPA, the calculation of the bremsstrahlung corrections is
based on the full lowest-order matrix elements to the processes
${\mathrm{e^+ e^-}} \to 4f+\gamma$.  This asymmetric treatment of
virtual and real photons requires a careful matching of the arising
infrared and collinear singularities.  We also take into account
higher-order initial-state photon radiation via the structure-function
method.  Here, we briefly describe the {\sc RacoonWW} approach, give
numerical results for the total W-pair production cross sections,
confront them with LEP2 data, and study the impact of the radiative
corrections on angular and W-invariant-mass distributions at LEP2 and
LC energies.
\end{abstract}

\section*{Introduction}
\setcounter{footnote}{0}
The precise measurements of the W-boson mass and the W-pair production
cross sections as well as the direct observation of triple-gauge-boson
couplings \cite{dw:lep2exp} at LEP2 provide further important
precision tests of the Standard Model of electroweak interactions.  To
match the experimental accuracy at LEP2 and at a future linear
collider (LC), the knowledge of the observed cross sections beyond
leading order in perturbation theory is crucial
\cite{dw:wwrev,dw:lep2rep,dw:lep2mcws}.  For instance, at LEP2 the
total W-pair production cross sections need to be known theoretically
to better than $1\%$ \cite{dw:lep2mcws}, which requires to go even
beyond universal radiative corrections such as the running of the
electromagnetic coupling, corrections connected to the $\rho$
parameter, the Coulomb singularity, and leading photonic corrections.

The full treatment of the processes ${\mathrm{e^+ e^-}} \to 4f$ at the
one-loop level is of enormous complexity \cite{dw:Vi98} and poses
severe gauge-invariance problems when introducing the finite W-boson
width. Fortunately, for W-pair production at LEP2 and at not too high
LC Center-of-Mass (CM) energies
\footnote{Above 0.5--$1\,$TeV at least the leading electroweak
logarithms at the two-loop level should be taken into account.}, it is
sufficient for the envisioned theoretical precision to take into
account only radiative corrections to those contributions that are
enhanced by two resonant W propagators.  The use of this so-called
double-pole approximation (DPA) for the calculation of the pair
production of unstable particles has been proposed in \cite{dw:Ae94}.
Recently different versions of the DPA for the calculation of
radiative corrections to off-shell W-pair production have been
described in the literature
\cite{dw:Be98,dw:yfsww,dw:yfswwnew,dw:ku99,dw:paper,dw:lep2}.

Here we present results of the first complete calculation of the
${\cal O}(\alpha)$ corrections for off-shell W-pair production in the
DPA that has been implemented in a Monte Carlo (MC) generator.  The
resulting MC program is called {\sc RacoonWW}, and numerical results
for the W-pair production cross sections and various distributions at
LEP2 and LC CM energies are provided in
\cite{dw:lep2mcws,dw:paper,dw:lep2} and \cite{dw:nlc}, respectively.
In \cite{dw:paper} we also presented a detailed description of the
{\sc RacoonWW} approach, discussed the intrinsic theoretical
uncertainty of our version of the DPA, and compared our results with
existing calculations (see also\cite{dw:lep2mcws}).

In the following sections we give a brief overview of the {\sc
RacoonWW} approach and provide numerical results for the total W-pair
production cross sections, angular and W-invariant-mass distributions
at the typical LEP2 and LC CM energies $\sqrt{s}=200\,$GeV and
500$\,$GeV, respectively.  For the same setup as described in
\cite{dw:lep2mcws}, we also present a comparison of {\sc RacoonWW}
results with the results of the Monte Carlo generator {\sc YFSWW3}
\cite{dw:yfsww,dw:yfswwnew} at $\sqrt{s}=500\,$GeV.

\section*{The {\sc RacoonWW} Approach}

\begin{figure}[b!] 
{\centerline{
\unitlength 1pt
\begin{picture}(120,90)(0,10)
\ArrowLine(30,50)( 5, 95)
\ArrowLine( 5, 5)(30, 50)
\Photon(30,50)(90,80){2}{6}
\Photon(30,50)(90,20){2}{6}
\GCirc(30,50){ 7}{0.5}
\Vertex(90,80){1.2}
\Vertex(90,20){1.2}
\ArrowLine(90,80)(120, 95)
\ArrowLine(120,65)(105,72.5)
\ArrowLine(105,72.5)(90,80)
\Vertex(105,72.5){1.2}
\ArrowLine(120, 5)( 90,20)
\ArrowLine( 90,20)(105,27.5)
\ArrowLine(105,27.5)(120,35)
\Vertex(105,27.5){1.2}
\Photon(105,27.5)(105,72.5){2}{4.5}
\put(93,47){$\gamma$}
\put(55,73){$W$}
\put(55,16){$W$}
\end{picture}}}
\vspace{10pt}
\caption[]{A typical diagram for virtual non-factorizable corrections
(the blob denotes the lowest-order W-pair production process).}
\label{F:dwnonf}
\end{figure}
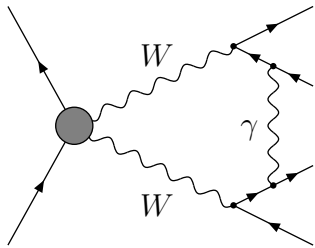

The radiative corrections to the processes ${\mathrm{e^+ e^- \to WW}}
\to 4f$ consist of virtual corrections as well as of real corrections
originating from the processes ${\mathrm{e^+ e^- \to WW}} \to 4f +
\gamma$.  In {\sc RacoonWW} only the virtual one-loop corrections are
treated in DPA, including the full set of {\it factorizable} and {\it
non-factorizable} ${\cal O}(\alpha)$ corrections to $ {\mathrm{e^+ e^-
\to WW}} \to 4f$ for massless fermions.  The {\it factorizable}
contributions comprise those virtual one-loop corrections that can
either be assigned to the W-pair production or to the W-decay
subprocesses.  The corresponding matrix element in DPA
\begin{equation}
{\cal M}_{\mathrm{DPA}}^{{\mathrm{e^+ e^-\to WW}}\to 4f}=
\frac{R(k_{\mathrm{W}^+}^2=M_{\mathrm W}^2,
k_{\mathrm{W}^-}^2=M_{\mathrm W}^2)}
{(k_{\mathrm{W}^+}^2 - M^2)(k_{\mathrm{W}^-}^2 - M^2)}
\end{equation}
is defined by the gauge-invariant residue $R(M_{\mathrm
W}^2,M_{\mathrm W}^2)$ and the W propagators with complex poles at
$M^2=M_{\mathrm W}^2-\mathrm{i} M_{\mathrm W} \Gamma_W$.  The one-loop
corrections to this residue can be deduced from the known results for
on-shell W-pair production \cite{dw:prod} and W~decay \cite{dw:wdec}.
The {\it non-factorizable} corrections \cite{dw:nonf} connect the
production and decay processes. In DPA, they comprise all
doubly-resonant contributions that are not already included in the
factorizable corrections.  Only those diagrams contribute where a
photon with energy $E_{\gamma}\lsim\Gamma_W$ is exchanged between the
production and decay subprocesses; all other non-factorizable diagrams
are negligible in DPA.  A typical diagram contributing to the
non-factorizable corrections is shown in Fig.~\ref{F:dwnonf}.
The calculation of the real ${\cal O}(\alpha)$ corrections is based on
the full lowest-order matrix element to ${\mathrm{e^+ e^-}} \to 4f +
\gamma$ for massless fermions as described in \cite{dw:photon}. In
this way, we avoid potential problems with a definition of a DPA for
real photons with energies $E_{\gamma} \sim \Gamma_W$.  However, since
we treat virtual and real photons differently, the former in DPA but
the latter completely, a careful matching of the arising infrared (IR)
and collinear singularities is needed: first the singularities are
extracted from the real photon contribution by using either a
subtraction \cite{dw:subtraction} or the phase-space-slicing method.
Both methods are implemented in {\sc RacoonWW}. As usual, these IR-
and collinear-singular contributions factorize into the Born cross
section and a factor containing the IR- and collinear-singular
logarithms, where the dependence on the photon phase space has been
partially integrated out analytically.  The remaining part of the real
corrections is free of singularities and can be evaluated with the
usual MC techniques.  Then, we extract those singular parts from the
IR- and collinear-singular contributions that exactly match the
singular parts of the virtual corrections, replace the full Born cross
section by the DPA Born cross section, and add this modified part to
the virtual corrections. This modification only introduces ambiguities
that are smaller than the accuracy of the DPA and leads to a proper
matching of IR and collinear singularities.
 
Beyond ${\cal O}(\alpha)$, {\sc RacoonWW} includes soft-photon
exponentiation and leading higher-order initial-state radiation (ISR)
up to ${\cal O}(\alpha^3)$ via the structure-function approach (see
e.g.\ \cite{dw:lep2rep} and references therein).  Also the
leading-order effects from $\Delta \rho$ and $\Delta \alpha$ are taken
into account by using the $G_{\mu}$ scheme.
 
By default, QCD corrections are taken into account by considering a
multiplicative factor $(1+\alpha_{\mathrm s}/ \pi)$ for each
hadronically decaying W boson. This affects the W-decay width in the W
propagators and the amplitudes to ${\mathrm{e^+e^-}} \to 4f$ for
quarks in the final state when the full phase space for gluons is
integrated over. In the total cross section these so-called ``naive''
QCD factors cancel.  If the gluon phase space is not integrated over,
the virtual and real QCD corrections to ${\mathrm{e^+ e^-}} \to 4f$
have to be calculated, which is technically similar to the calculation
of the photonic corrections.  This option is also supported by {\sc
RacoonWW}.
 
\section*{Numerical Results}
For the numerical results we use the following parameters:
\begin{equation}
\begin{array}[b]{rclrcl}
G_{\mu} &=& 1.16637\times 10^{-5} \, \mathrm{GeV}^{-2}, 
\qquad&\alpha&=&1/137.0359895, \\
M_{\mathrm W} &=& 80.35 \, \mathrm{GeV},& 
\Gamma_{\mathrm W} &=& 2.08699\ldots \mathrm{GeV}, \\
M_{\mathrm Z} &=& 91.1867 \, \mathrm{GeV},& 
\Gamma_{\mathrm Z} &=& 2.49471 \, \mathrm{GeV}, \\
m_{\mathrm t} &=& 174.17 \, \mathrm{GeV},& 
M_{\mathrm H}&=& 150 \, \mathrm{GeV}, \\
m_{\mathrm e} &=& 510.99907 \, \mathrm{keV},
& \alpha_{\mathrm s}&=& 0.119.
\end{array}
\end{equation}
We work in the fixed-width scheme and fix the weak mixing angle by
using $c_{\mathrm w}=M_{\mathrm W}/M_{\mathrm Z}$, $s_{\mathrm
w}^2=1-c_{\mathrm w}^2$.  These parameters are over-complete but
self-consistent. Instead of $\alpha$ we use $G_{\mu}$ to parametrize
the lowest-order matrix element, i.e.\ we use the effective coupling
\begin{equation}
\alpha_{G_{\mu}} = \frac{\sqrt{2} G_{\mu} M_{\mathrm W}^2 s_{\mathrm w}^2}{\pi}
\end{equation}
in the lowest-order matrix element. This parameterization has the
advantage that all higher-order contributions associated with the
running of the electromagnetic coupling from $q^2=0$ to
$q^2=M_{\mathrm W}^2$ and the leading universal two-loop $m_{\mathrm
t}$-dependent corrections are already absorbed in the Born cross
section.  In the relative ${\cal O}(\alpha)$ corrections, on the other
hand, we use $\alpha=\alpha(0)$, since in the real corrections, which
yield the bulk of the remaining corrections, the scale of the real
photon is zero.  The W-boson width given above is calculated including
the electroweak and QCD one-loop corrections.

In the following, for the lowest-order contributions only the CC03
diagrams are taken into account. Our ``best'' results comprise the
CC03 Born cross sections and the corrected cross sections including
all the radiative corrections briefly described in the previous
section (more details can be found in \cite{dw:paper}), i.e.\ the
electroweak one-loop corrections in DPA, the exact ${\cal O}(\alpha)$
photon radiation based on the full matrix element for
${\mathrm{e^+e^-}} \to 4f \gamma$, higher-order ISR up to ${\cal
O}(\alpha^3)$, and ``naive'' QCD ${\cal O}(\alpha_{\mathrm{s}})$
corrections.

\begin{figure}[b!] 
\centerline{\epsfig{file=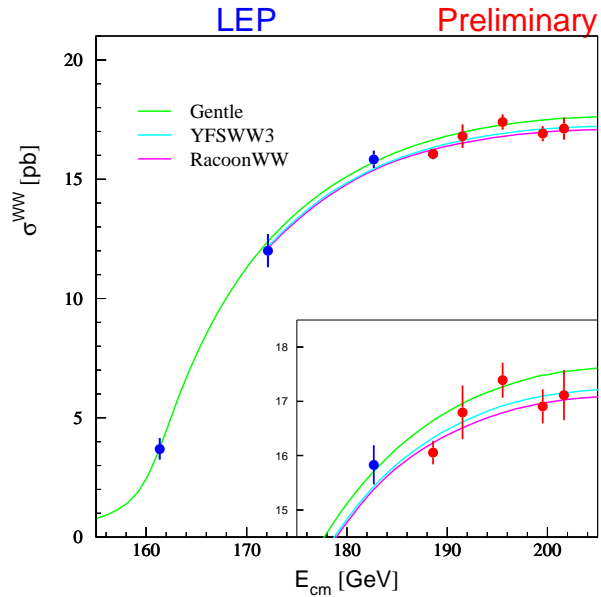,height=3.5in,width=3.5in}}
\vspace{10pt}
\caption[]{Total W-pair production cross section at LEP2, as given by the
LEPEWWG \cite{dw:LEPEWWG}.}
\label{F:dwackeroth:1}
\end{figure}

Figure~\ref{F:dwackeroth:1} shows a comparison of the results of {\sc
RacoonWW} for the total CC03 W-pair production cross section (see also
\cite{dw:lep2mcws}) and of other Monte Carlo predictions with recent
LEP2 data, as given by the LEP Electroweak Working Group
\cite{dw:LEPEWWG} for the Winter 2000 conferences.
The data are in good agreement with the predictions of {\sc RacoonWW}
and {\sc YFSWW3} \cite{dw:yfsww,dw:yfswwnew}.  At the time of the
conferences in Winter 2000 the predictions of {\sc YFSWW3} were about
$0.5$--$0.7\%$ larger than those of {\sc RacoonWW}, which is somewhat
larger than the intrinsic DPA ambiguity. Meanwhile, however, the main
source of this discrepancy was found \cite{dw:lep2mcws}, and the new
{\sc YFSWW3} \cite{dw:yfswwnew} results differ from the ones of {\sc
RacoonWW} only by about $0.3\%$ at LEP2 CM energies.  More details on
the conceptual differences of the two generators, as well as a
detailed comparison of numerical results for LEP2 energies, can be
found in \cite{dw:lep2mcws} and below.  Figure~\ref{F:dwackeroth:1}
also includes the prediction provided by {\sc GENTLE}
\cite{dw:gentle}, which is 2--2.5\%\ larger than those from {\sc
RacoonWW} and {\sc YFSWW3}.  This difference is due to the neglect of
non-leading, non-universal ${\cal O}(\alpha)$ corrections in {\sc
GENTLE}.  In summary, the comparison between SM predictions and LEP2
data reveals evidence of non-leading electroweak radiative corrections
beyond the level of universal effects.

In the following we study the impact of the radiative corrections on
the W-invariant-mass distributions and angular distributions and use
the separation and recombination cuts of \cite{dw:lep2mcws,dw:paper}:
\begin{enumerate}
\item All photons within a cone of 5 degrees around the beams are
  treated as invisible, i.e.\ their momenta are disregarded when
  calculating angles, energies, and invariant masses.
\item Next, the invariant masses 
  $M_{f\gamma}$ of the photon with
  each of the charged final-state fermions are calculated. If the
  smallest one is smaller than 
  a certain cutoff $M_{\mathrm{rec}}$ or if the energy 
  of the photon is smaller than 1$\,$GeV, 
  the photon is combined with the charged
  final-state fermion that leads to the smallest $M_{f\gamma}$,
  i.e.\ the momenta of the photon and the
  fermion are added and associated with the momentum of the fermion,
  and the photon is discarded.
\item Finally, all events are discarded
  in which one of the charged
  final-state fermions is within a cone of 10 degrees around the
  beams.  No other cuts are applied.
\end{enumerate}
We consider two recombination cuts: $M_{\mathrm rec}=5\,$GeV ({\it bare}) 
and $M_{\mathrm rec}=25\,$GeV ({\it calo}).

\begin{figure}[b!] 
{\centerline{
\setlength{\unitlength}{1cm}
\begin{picture}(14,6.8)
\put(-4.2,-13.5){\includegraphics{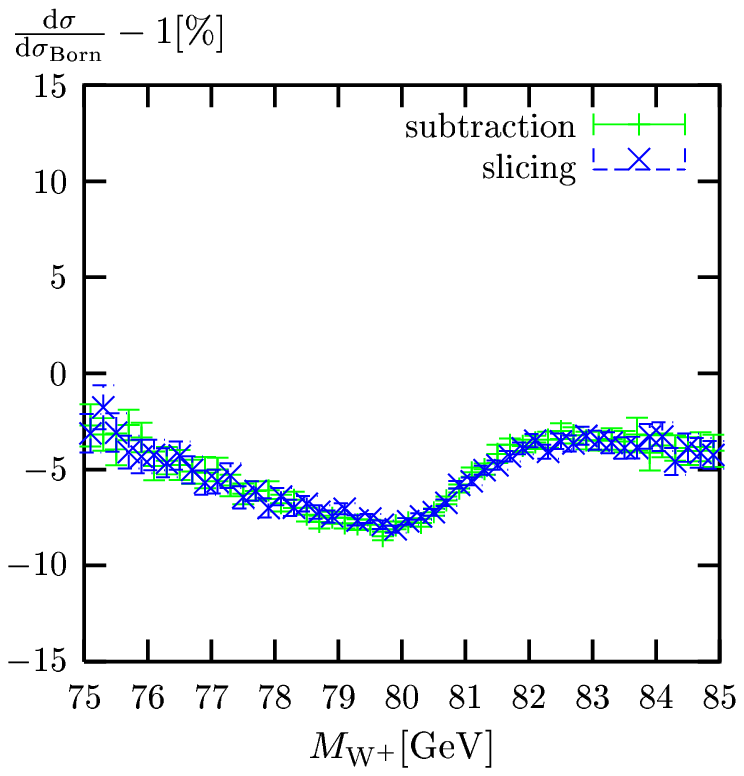}}
\put( 2.8,-13.5){\includegraphics{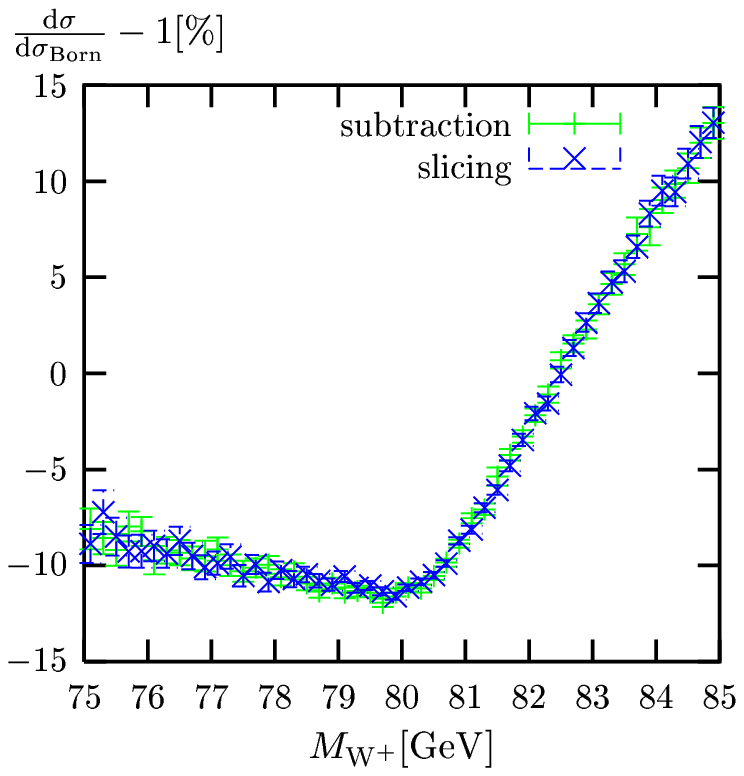}}
\put( 1.4,5.0){\footnotesize $M_{\mathrm rec}=5\,$GeV}
\put( 8.4,5.0){\footnotesize $M_{\mathrm rec}=25\,$GeV}
\end{picture} }}
\vspace{10pt}
\caption[]{Relative corrections to the $\mathrm{W}^+$ invariant-mass
distributions for ${\mathrm{e^+ e^- \to u \bar d}} \mu^-
\bar\nu_{\mu}$ at $\sqrt{s}=200\,$GeV when {\it bare} (l.h.s) or {\it
calo} (r.h.s) cuts are applied (taken from \cite{dw:paper}).}
\label{F:dwackeroth:2}
\end{figure}

\begin{figure}[b!] 
{\centerline{
\setlength{\unitlength}{1cm}
\begin{picture}(14,6.8)
\put(-4.2,-13.5){\includegraphics{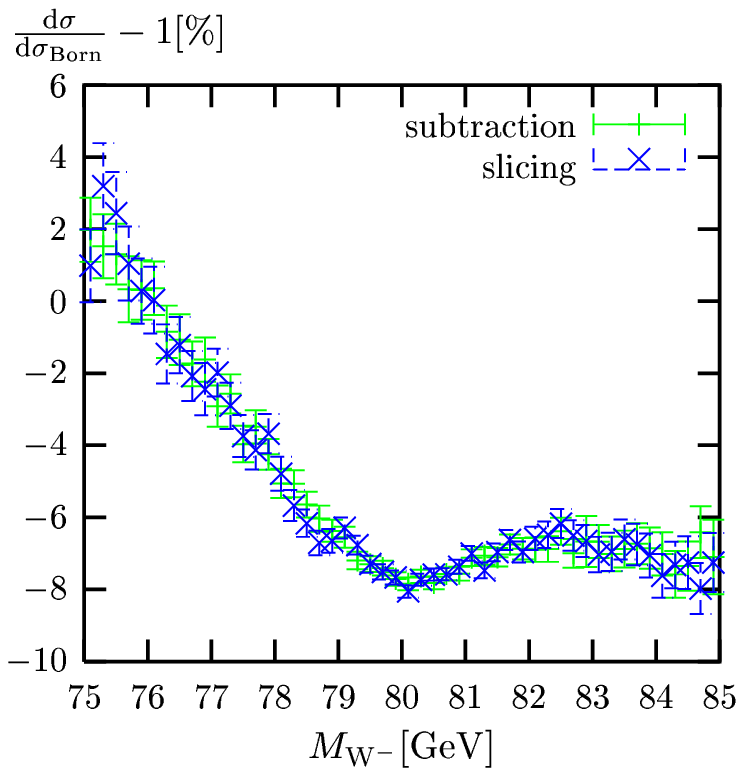}}
\put( 2.8,-13.5){\includegraphics{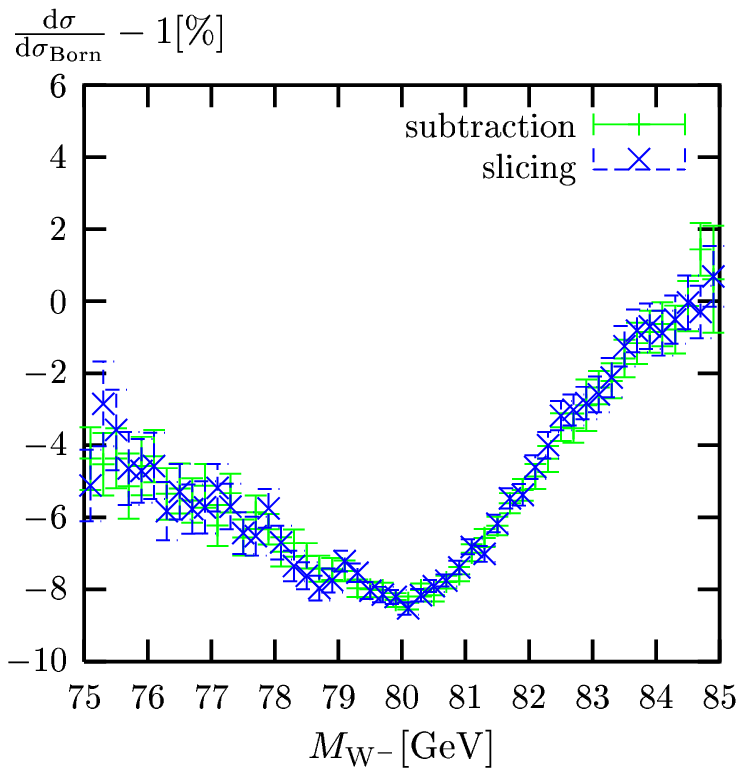}}
\put( 1.4,5.0){\footnotesize $M_{\mathrm rec}=5\,$GeV}
\put( 8.4,5.0){\footnotesize $M_{\mathrm rec}=25\,$GeV}
\end{picture} }}
\vspace{10pt}
\caption[]{Relative corrections to the $\mathrm{W}^-$ invariant-mass
distributions for ${\mathrm{e^+ e^- \to u \bar d}} \mu^-
\bar\nu_{\mu}$ at $\sqrt{s}=200\,$GeV when {\it bare} (l.h.s) or {\it
calo} (r.h.s) cuts are applied (taken from \cite{dw:paper}).}
\label{F:dwackeroth:3}
\end{figure}

\begin{figure}[t!] 
{\centerline{
\setlength{\unitlength}{1cm}
\begin{picture}(14,6.8)
\put(-4.2,-13.5){\includegraphics{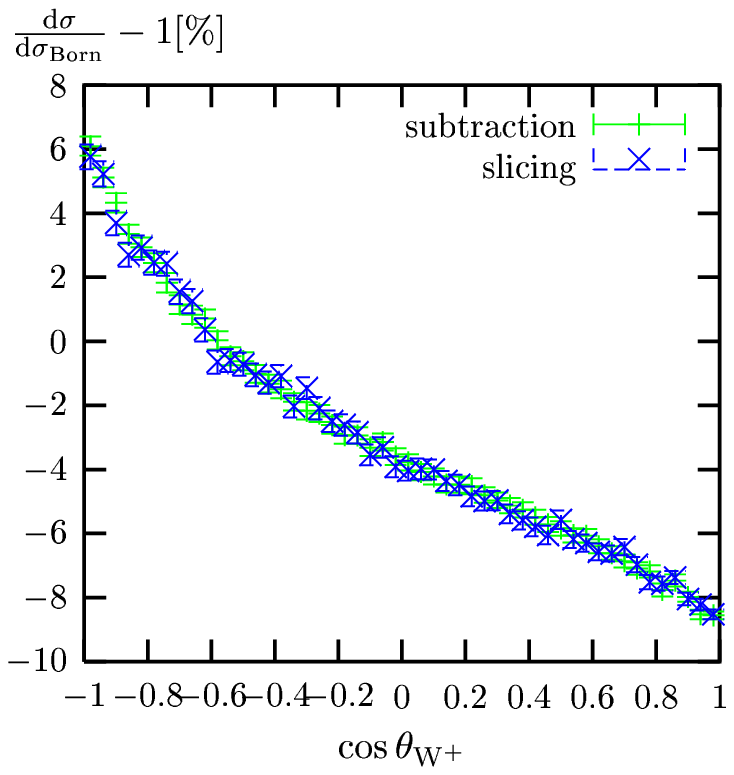}}
\put( 2.8,-13.5){\includegraphics{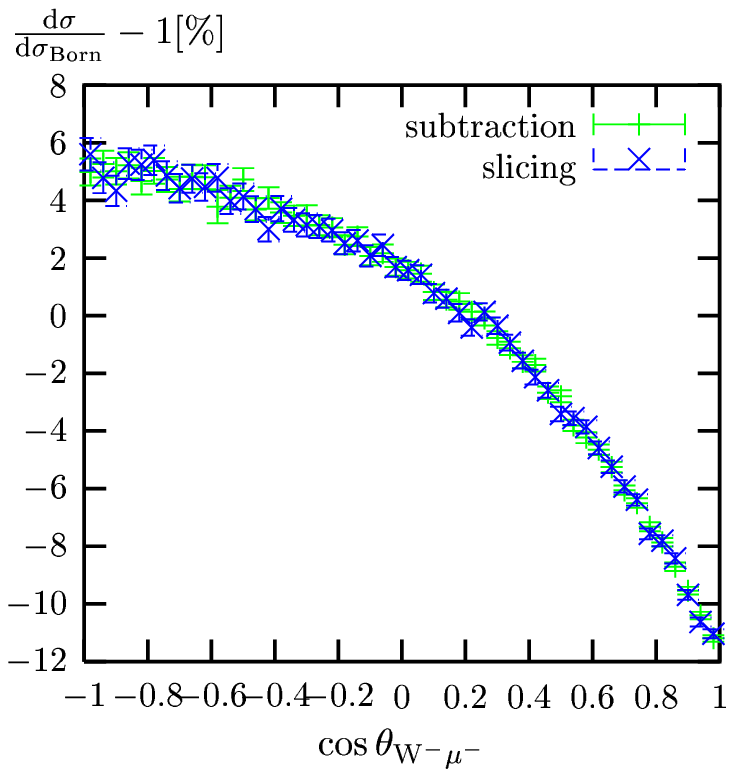}}
\end{picture} }}
\vspace{10pt}
\caption[]{Relative corrections to the distributions in the cosines of
the $\mathrm{W}^+$ production angle (l.h.s) and the $\mu^-$ scattering
angle with respect to the $\mathrm{W}^-$ direction (r.h.s.)  for
${\mathrm{e^+ e^- \to u \bar d}} \mu^- \bar\nu_{\mu}$ at
$\sqrt{s}=200\,$GeV when {\it bare} cuts are applied.}
\label{F:dwackeroth:4}
\end{figure}

As mentioned before, {\sc RacoonWW} contains two branches for the
treatment of IR and collinear singularities, one following the
subtraction method of \cite{dw:subtraction} and one the
phase-space-slicing method.  While these two branches use the same
matrix elements, the Monte Carlo integration is performed completely
independently, thus providing us with a powerful numerical check of
{\sc RacoonWW}. In the following we present numerical results of both
branches of {\sc RacoonWW}.  In Table \ref{T:dwackeroth:1} we provide
the CC03 Born and ``best'' results for the total W-pair production
cross sections for the semileptonic process ${\mathrm{e^+ e^- \to u
\bar d}} \mu^- \bar\nu_{\mu}(\gamma)$ at the CM energies
$\sqrt{s}=200\,$GeV and 500$\,$GeV with the {\it bare} or {\it calo}
cuts applied.  The results obtained when using the phase-space-slicing
method agree well with those of the subtraction method, i.e.\ within
the statistical errors of $0.05$--$0.09\%$.
\begin{table}
\caption{Born and ``best'' predictions for the total CC03
cross-sections for ${\mathrm{e^+ e^- \to u \bar d}} \mu^-
\bar\nu_{\mu}(\gamma)$ when using the subtraction method or the
phase-space-slicing method at $\sqrt{s}=$ 200$\,$GeV and 500$\,$GeV
when {\it bare} or {\it calo} cuts are applied.  The numbers in
parentheses are statistical errors corresponding to the last digits.}
\label{T:dwackeroth:1}
\begin{tabular}{|c|c|c|c||c|c|c|} \hline
& \multicolumn{3}{|c||}{${ \sqrt{s}=200 \; \mathrm{GeV}}$}
& \multicolumn{3}{|c|}{${ \sqrt{s}=500 \; \mathrm{GeV}}$}\\ \cline{2-7}
${\bf \sigma_{\mathrm{tot}}}$ [fb] 
& & \multicolumn{2}{|c||}{best} & & \multicolumn{2}{|c|}{best} \\ \cline{3-4} 
\cline{6-7}
 & \hspace*{0.05cm}  \raisebox{1.5ex}[-1.5ex]{Born} \hspace*{0.05cm} 
 & \hspace*{0.05cm} {\it bare} \hspace*{0.05cm}  
 & \hspace*{0.5cm} {\it calo} \hspace*{0.5cm} 
 & \hspace*{0.05cm}  \raisebox{1.5ex}[-1.5ex]{Born} \hspace*{0.05cm} 
 & \hspace*{0.05cm} {\it bare} \hspace*{0.05cm}   
 & \hspace*{0.05cm} {\it calo} \hspace*{0.05cm}   \\ \hline \hline
slicing     & 627.38(11)   & 591.11(25)   & 590.92(25) 
            & 181.46(6)    & 198.79(15)   & 198.60(15) \\
subtraction & 627.22(12)   & 590.94(14)   & 590.82(14) 
            & 181.51(6)    & 198.70(8)\phantom{0}   
            & 198.52(8)\phantom{0}\\ \hline 
(sub--sli)/sli & $-0.03(3)\%$ & $-0.03(5)\%$ & $-0.02(5)\%$ 
               & $0.03(5)\%$  & $-0.05(9)\%$ & $-0.04(9) \%$ \\ 
\hline\hline
\end{tabular}
\end{table}
In Figs.~\ref{F:dwackeroth:2}, \ref{F:dwackeroth:3}, and
\ref{F:dwackeroth:4} we show the relative corrections
$\mathrm{d}\sigma/ \mathrm{d}\sigma_{\mathrm{Born}}-1$ (in per cent),
where $\mathrm{d}\sigma_{\mathrm{Born}}$ and $\mathrm{d}\sigma$ denote
the CC03 Born and ``best'' predictions, respectively, for a variety of
distributions for ${\mathrm{e^+ e^- \to u \bar d}} \mu^-
\bar\nu_{\mu}(\gamma)$ at the LEP2 CM energy $\sqrt{s}=$200$\,$GeV.
The relative corrections to the W-invariant-mass distributions are
shown in Figs.~\ref{F:dwackeroth:2} and \ref{F:dwackeroth:3} for the
{\it bare} and {\it calo} recombination cuts.  The invariant masses
are obtained from the four-momenta of the decay fermions of the W
bosons after eventual recombination with the photon
four-momentum. They are particularly sensitive to the treatment of the
photons, and thus a strong dependence of the relative corrections on
the recombination cut can be observed. The observed distortion of the
W-invariant-mass distributions is of particular interest when the
W-boson mass is reconstructed from the W-decay products.  A detailed
discussion of the distortion of the invariant-mass distributions of
the W bosons and their origin has been given in \cite{dw:lep2}.  The
distributions obtained from the phase-space-slicing and subtraction
methods are compatible with each other. The integration errors are
larger further away from the W resonance because of the decreasing
statistics.
For the angular distributions, we define all angles in the laboratory
system, which is the CM system of the initial state.  The relative
corrections to the distributions in the cosines of the $\mathrm{W}^+$
production angle $\theta_{W^+}$ and of the decay angle $\theta_{W^-
\mu^-}$ are shown in Fig.~\ref{F:dwackeroth:4} when {\it bare} cuts
are applied. In contrast to the invariant-mass distributions, the
angular distributions hardly depend on the recombination procedure
apart from very large angles where the cross section is small. Thus,
similar results are obtained when the {\it calo} recombination cut is
used \cite{dw:paper}.  The distributions obtained with the subtraction
and the phase-space-slicing branches of {\sc RacoonWW} agree with each
other within the integration errors, i.e.\ in general to better than
$0.5\%$.  The increase of the integration errors for large decay
angles is due to the smallness of the corresponding cross section.

\subsubsection*{Comparison with {\sc YFSWW3} at $\mathrm{\sqrt{s}=500\,GeV}$}

In \cite{dw:lep2mcws} a tuned comparison of {\sc RacoonWW} and {\sc
YFSWW3} of predictions for ${\mathrm{e^+ e^-\to WW}} \to 4f(\gamma)$
observables at LEP2 CM energies has been performed.  This comparison
together with a study of the intrinsic uncertainty of the DPA with
{\sc RacoonWW}, as well as a comparison with a semi-analytical
calculation \cite{dw:Be98}, enabled the assignment of an overall
theoretical uncertainty of $0.4\%$ to the state-of-the-art MC
predictions for the total W-pair production cross sections at
200$\,$GeV \cite{dw:lep2mcws}. This confirms the naively expected DPA
accuracy of about 0.5\% for $\sqrt{s}\gsim180\,$GeV for the total CC03
cross sections.  In \cite{dw:paper} we studied the intrinsic
ambiguities of the ${\cal O}(\alpha)$ corrections in DPA also at
500$\,$GeV and found them to be 0.1\% for the total W-pair production
cross section and a few per mil whenever W-pair production dominates
the considered observable. A comparison of {\sc RacoonWW} and {\sc
YFSWW3} results for the total W-pair production cross sections at
200$\,$GeV and 500$\,$GeV, when no or bare cuts are applied, is given
in Table \ref{T:dwackeroth:2}.  The relative differences are at the
$0.3\%$ level, which is of the order of the intrinsic ambiguity of any
DPA implementation.
\begin{table}
\caption[]{Born and ``best'' predictions for the total CC03
cross-sections for ${\mathrm{e^+ e^- \to u \bar d}} \mu^-
\bar\nu_{\mu}(\gamma)$ obtained with {\sc RacoonWW}
(subtraction-method branch) and {\sc YFSWW3} (scheme A)
\cite{dw:yfswwnew} at $\sqrt{s}=$ 200$\,$GeV and 500$\,$GeV when no or
{\it bare} cuts are applied.  The numbers in parentheses are
statistical errors corresponding to the last digits.}
\label{T:dwackeroth:2}
\begin{tabular}{|c|c|c||c|c|} \hline
{\bf no cuts} & \multicolumn{2}{|c||}{${ \sqrt{s}=200 \; \mathrm{GeV}}$}
& \multicolumn{2}{|c|}{${ \sqrt{s}=500 \; \mathrm{GeV}}$}\\ \cline{2-5}
${\bf \sigma_{\mathrm{tot}}}$ [fb] 
& Born & best & Born & best \\ \hline \hline
{\sc YFSWW3}   & 659.64(07)  & 622.71(19)  & 261.377(34) & 279.086(97) \\
{\sc RacoonWW} & 659.51(12)  & 621.06(14)  & 261.400(70) & 280.149(86) 
\\ \hline 
(Y--R)/Y       & $0.02(2)\%$ & $0.27(4)\%$ & $-0.01(3)\%$ & $-0.38(5)\%$ \\ 
\hline\hline
{\bf {\it bare}  cuts} 
& \multicolumn{2}{|c||}{${ \sqrt{s}=200 \; \mathrm{GeV}}$}
& \multicolumn{2}{|c|}{${ \sqrt{s}=500 \; \mathrm{GeV}}$}\\ \cline{2-5}
${\bf \sigma_{\mathrm{tot}}}$ [fb] 
& Born & best & Born & best \\ \hline \hline
{\sc YFSWW3}   & 627.18(07)   & 592.68(19)  & 181.507(33) & 197.933(84) \\
{\sc RacoonWW} & 627.22(12)   & 590.94(14)  & 181.507(63) & 198.696(76) \\
 \hline 
(Y--R)/Y & $-0.01(2)\%$ & $0.29(4)\%$ & $0.00(4)\%$ & $-0.39(6)\%$ \\ 
\hline\hline
\end{tabular}
\end{table}

\begin{figure}[b!] 
{\centerline{
\setlength{\unitlength}{1cm}
\begin{picture}(14,6.8)
\put(-4.2,-13.5){\includegraphics{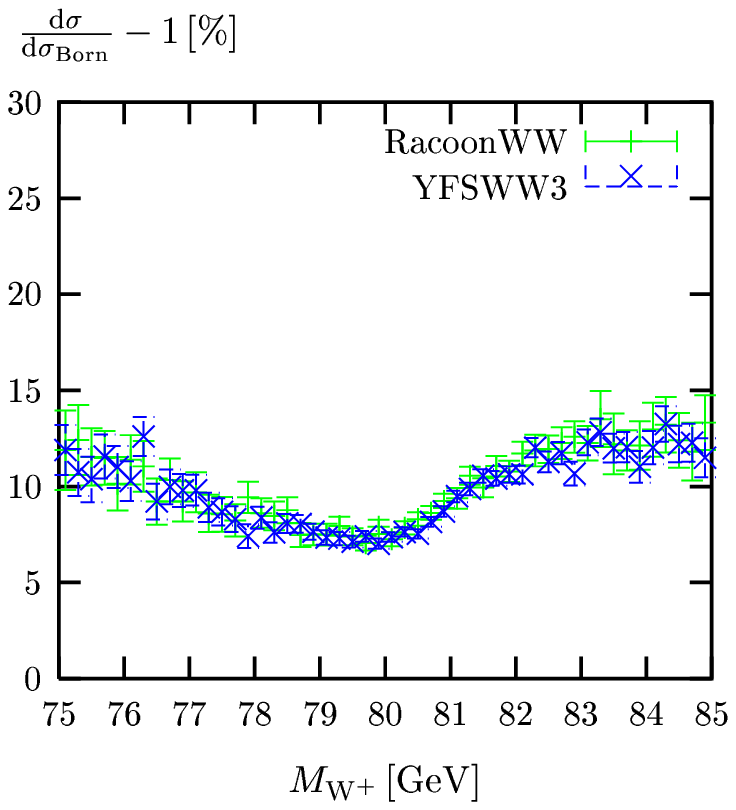}}
\put( 2.8,-13.5){\includegraphics{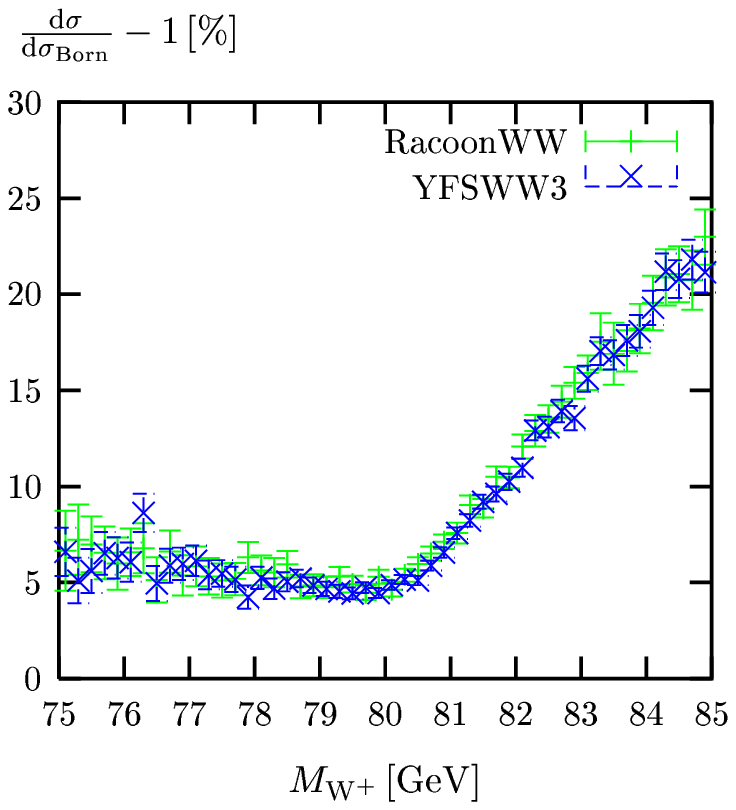}}
\put( 1.4,5.0){\footnotesize $M_{\mathrm rec}=5\,$GeV}
\put( 8.4,5.0){\footnotesize $M_{\mathrm rec}=25\,$GeV}
\end{picture} }}
\vspace{10pt}
\caption[]{Relative corrections to the $\mathrm{W}^+$ invariant-mass
distributions for ${\mathrm{e^+ e^- \to u \bar d}} \mu^- \bar\nu_{\mu}$
at $\sqrt{s}=500\,$GeV when {\it bare} (l.h.s) or {\it calo} (r.h.s)
cuts are applied.}
\label{F:dwackeroth:5}
\end{figure}

\begin{figure}[b!] 
{\centerline{
\setlength{\unitlength}{1cm}
\begin{picture}(14,6.8)
\put(-4.2,-13.5){\includegraphics{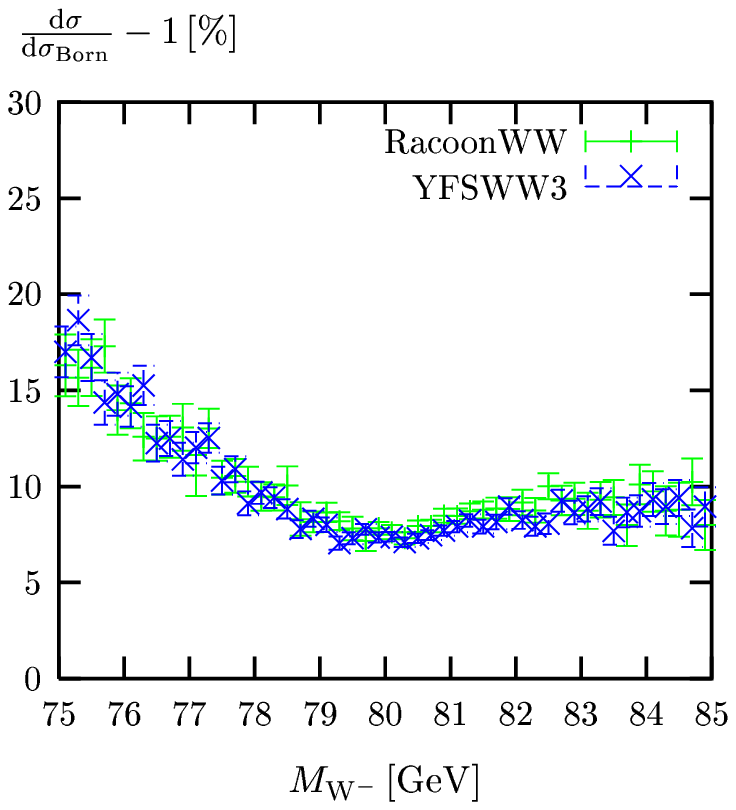}}
\put( 2.8,-13.5){\includegraphics{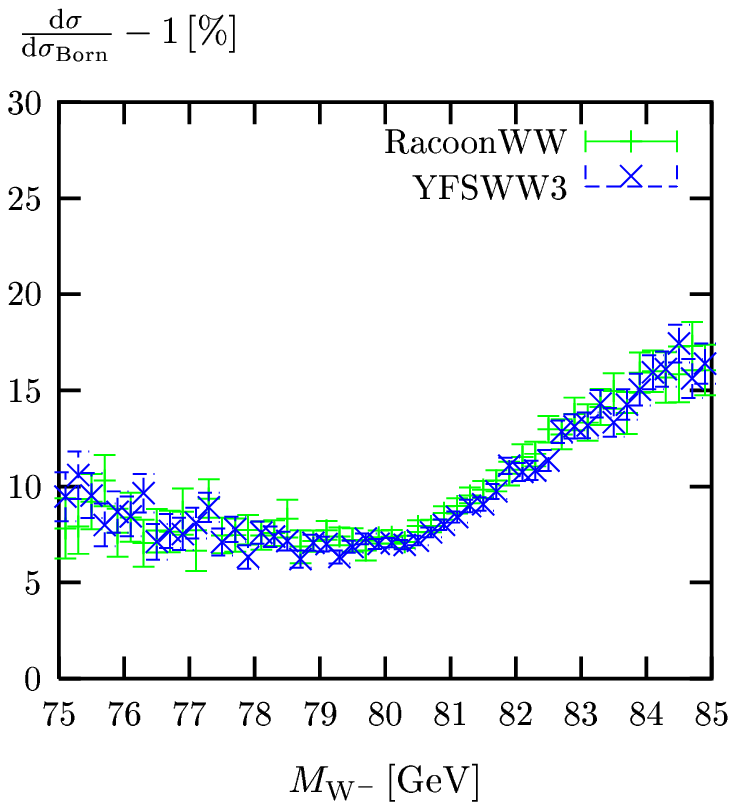}}
\put( 1.4,5.0){\footnotesize $M_{\mathrm rec}=5\,$GeV}
\put( 8.4,5.0){\footnotesize $M_{\mathrm rec}=25\,$GeV}
\end{picture} }}
\vspace{10pt}
\caption[]{Relative corrections to the $\mathrm{W}^-$ invariant-mass
distributions for ${\mathrm{e^+ e^- \to u \bar d}} \mu^- \bar\nu_{\mu}$
at $\sqrt{s}=500\,$GeV when {\it bare} (l.h.s) or {\it calo} (r.h.s)
cuts are applied.}
\label{F:dwackeroth:6}
\end{figure}

\begin{figure}[b!] 
{\centerline{
\setlength{\unitlength}{1cm}
\begin{picture}(14,6.8)
\put(-4.2,-13.5){\includegraphics{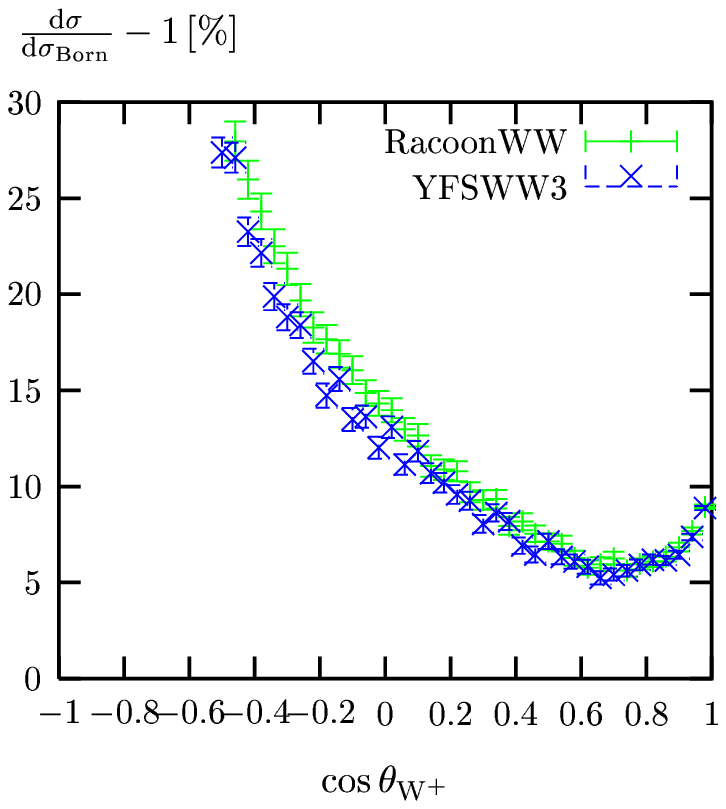}}
\put( 2.8,-13.5){\includegraphics{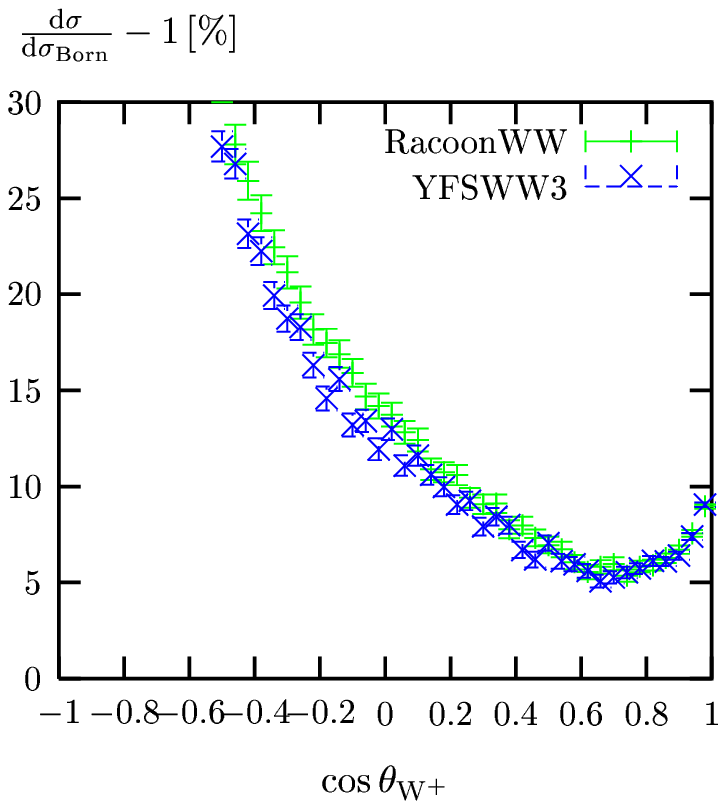}}
\put( 1.4,1.5){\footnotesize $M_{\mathrm rec}=5\,$GeV}
\put( 8.4,1.5){\footnotesize $M_{\mathrm rec}=25\,$GeV}
\end{picture} }}
\vspace{10pt}
\caption[]{Relative corrections to the cosine of the 
  $\mathrm{W}^+$-production-angle distributions for ${\mathrm{e^+ e^-
      \to u \bar d}} \mu^- \bar\nu_{\mu}$ at $\sqrt{s}=500\,$GeV when
  {\it bare} (l.h.s) or {\it calo} (r.h.s) cuts are applied.}
\label{F:dwackeroth:7}
\end{figure}

In Figs.~\ref{F:dwackeroth:5}, \ref{F:dwackeroth:6}, and
\ref{F:dwackeroth:7} we show the predictions of the subtraction-method 
branch of {\sc RacoonWW} together with the predictions of {\sc
YFSWW3} (scheme A) \cite{dw:yfswwnew} for the relative corrections to
the W-invariant-mass and $\cos\theta_{\mathrm{W}^+}$ distributions to
${\mathrm{e^+ e^- \to u \bar d}} \mu^- \bar\nu_{\mu} (\gamma)$ at 
500$\,$GeV when {\it bare} or {\it calo} cuts are applied.
In contrast to the LEP2 case, the radiative corrections now mostly
increase the Born cross sections and are especially pronounced in the
angular distributions.  The dramatic increase of the relative
corrections to the $\mathrm{W}^+$-production-angle distribution of
Fig.~\ref{F:dwackeroth:7} is due to initial-state hard photon
radiation which causes a redistribution of events to a phase-space
region where the Born cross sections are very small.  A detailed
discussion of the invariant-mass and angular distributions obtained
with {\sc RacoonWW} at 500$\,$GeV is given in \cite{dw:nlc}.

The invariant-mass distributions in Figs.~\ref{F:dwackeroth:5} and
\ref{F:dwackeroth:6} obtained by {\sc RacoonWW} and {\sc YFSWW3} are
statistically compatible with each other for both recombination
procedures, i.e.\ they agree to better than 1\%.  The predictions of
{\sc RacoonWW} and {\sc YFSWW3} for the distributions in the cosine of
the $\mathrm{W}^+$ production angle, shown in
Fig.~\ref{F:dwackeroth:7}, agree within the statistical errors for
small angles but differ by up to several per cent for intermediate and
large angles where the cross sections become small. This behaviour is
independent of the applied recombination cuts.
\section*{Acknowledgments}
We thank the authors of {\sc YFSWW3}, S. Jadach, W. P\l{}aczek,
M. Skrzypek, B.F.L. Ward, and Z. W\c{a}s, for helpful discussions
about their results and for making them available to us before
publication.

\end{document}